# Thermoelectric bolometers based on ultra-thin heavily doped single-crystal silicon membranes

Andrey V. Timofeev [a], Aapo Varpula [a], Andrey Shchepetov, Kestutis Grigoras, Juha Hassel, Jouni Ahopelto, Markku Ylilammi, and Mika Prunnila [b]

VTT Technical Research Centre of Finland Ltd, P.O. Box 1000, FI-02044, Espoo, Finland

[a] *These authors have contributed equally to the work*

[b] *mika.prunnila@vtt.fi*

**Abstract** - We present ultra-thin silicon membrane thermocouple bolometers suitable for fast and sensitive detection of low levels of thermal power and infrared radiation at room temperature. The devices are based on 40 nm-thick strain tuned single crystalline silicon membranes shaped into heater/absorber area and narrow n- and p-doped beams, which operate as the thermocouple. The electro-thermal characterization of the devices reveal noise equivalent power of 13 pW/Hz$^{1/2}$ and thermal time constant of 2.5 ms. The high sensitivity of the devices is due to the high Seebeck coefficient of 0.39 mV/K and reduction of thermal conductivity of the Si beams from the bulk value. The bolometers operate in the Johnson-Nyquist noise limit of the thermocouple, and the performance improvement towards the operation close to the temperature fluctuation limit is discussed.

Bolometers are thermal detectors where energy flux (typically from photons) heats or cools an absorber leading to a measurable electrical signal caused by the flux induced temperature change. In contrast to photodetectors based on photo-carrier generation in semiconductors, bolometers provide a cost effective method, e.g., for the detection of infrared (IR) radiation, which has numerous applications ranging from chemical sensors to thermal imagers [1, 2]. The un-cooled room temperature bolometers utilize various effects to detect the induced temperature change. These include, for example, the temperature dependence of resistivity (resistance) or permittivity (capacitance) and the pyroelectric and the thermoelectric effects [3, 2].

During the last decades nano-thermoelectrics has emerged as a new research field [4, 5]. Nano-thermoelectrics utilizes different combinations of nano-fabrication and material synthesis techniques to engineer charge carrier and/or phonon transport in order to enhance thermoelectric performance, while the traditional thermoelectrics relies more on the bulk material properties. Scaling down the cross-sectional dimensions of the thermoelectric electrodes (or using a bottom-up approach with nanowires) can lead to strong reduction in the phonon heat conduction while essentially preserving the thermoelectric power factor and this approach has been recognized as highly effective method to enhance the performance of thermoelectric devices. Here, even silicon, which in bulk form has an extremely high phonon thermal conductivity, has been demonstrated to have attractive thermoelectric performance when scaled down to the sub-100 nm regime [6, 7].

Nano-thermoelectrics has been mainly driven by thermal energy harvesting with less attention towards thermoelectric bolometers (often referred to as thermopiles or thermocouples). In this work, we demonstrate an un-cooled high sensitivity thermoelectric bolometer based on suspended 40 nm-thick single-crystalline silicon thermocouple. In addition, we discuss how the performance of such devices can be further enhanced and even brought close to the ultimate sensitivity limit of thermal detectors determined by the thermal fluctuation noise. Our approach is based on strain tuned Si membranes [8] used previously as a (passive) platform, for example, in the studies of strain tuning of the group velocity of 2D phonons [9] and engineering of phonon surface scattering [10]. In addition of IR detection the demonstrated bolometer devices can be also used as a heat flux sensing platform in scanning thermal microscopy [11] and near-field heat transfer experiments [12].

The distinct feature of our devices is the absence of any kind of passive micromechanical support layer (e.g. $Si_3N_4$) which is commonly used in micromachined bolometers. The free-standing beams of heavily doped n- and p- silicon with corresponding thermoelectric coefficients $S_n$ and $S_p$ (subscript *n* refers to n-type material and subscript *p* to p-type material in this work) and resistances $R_n$ and $R_p$ form the thermocouple element and the support at the same time (Fig. 1). This approach minimizes the total thermal conductance *G* of the suspended device to the substrate and thereby enhances the bolometer responsivity $\frac{dV}{dP} = \frac{dV}{GdT}\frac{1}{\sqrt{1+(2\pi f \tau_0)^2}}$ to the incident thermal power *P* at frequency *f*. Here, *V* is the thermoelectric voltage induced by the temperature gradient $\Delta T$ between the membrane and the thermal bath, and $S = \frac{dV}{dT} = S_p - S_n$ is the total Seebeck coefficient of the thermocouple. The lack of extra support also reduces the heat capacity *C* which often limits thermal time constant $\tau_0$ = C/G. The key figure of merit of a bolometer is the power resolution per a unit frequency bandwidth given by the noise equivalent power **NEP(*f*)**. It can be expressed as the sum of the two dominant uncorrelated contributions given by the phonon noise due to the temperature fluctuations across the thermal link *G* giving contribution $\mathbf{NEP}_{\text{ph}} = \sqrt{4k_B T^2 G}$, and Johnson noise of the total thermocouple resistance $R = R_n + R_p$ giving contribution $\mathbf{NEP}_J(f) = \left(\frac{dV}{dP}\right)^{-1}\sqrt{4k_B TR}$. The square of the total NEP thus reads

$$\mathbf{NEP}^2(f) = \mathbf{NEP}_{\text{ph}}^2 + \mathbf{NEP}_J^2(f) = \mathbf{NEP}_{\text{ph}}^2\left(1 + \frac{1+(2\pi f \tau_0)^2}{\widetilde{ZT}}\right), \quad (1)$$

where $\widetilde{ZT} = S^2 T/(GR)$ is the effective thermoelectric figure of merit of bolometer. To maximize $\widetilde{ZT}$ at fixed operating temperature $T$, in addition to the reduction of $G$ the ratio $S^2/R$ term needs to be maximized, which can be achieved in practice by the optimal doping level at high carrier concentration limit of ~$10^{19}$-$10^{20}$ cm$^{-3}$ [13, 14, 15]. In the case where $G$ and $R$ are entirely dominated by the thermal conductance and electrical resistance of the thermocouple beams, respectively, $\widetilde{ZT}$ coincides with the ideal thermoelectric figure of merit $ZT = \frac{S^2 T}{\left(\frac{k_n w_n}{l_n}+\frac{k_p w_p}{l_p}\right)\left(\frac{\rho_n l_n}{w_n}+\frac{\rho_p l_p}{w_p}\right)}$, where $k_n$ and $k_p$ are the thermal conductivities, and $\rho_n$ and $\rho_p$ electrical resistivities of the of the n- and p- legs of the same thickness.

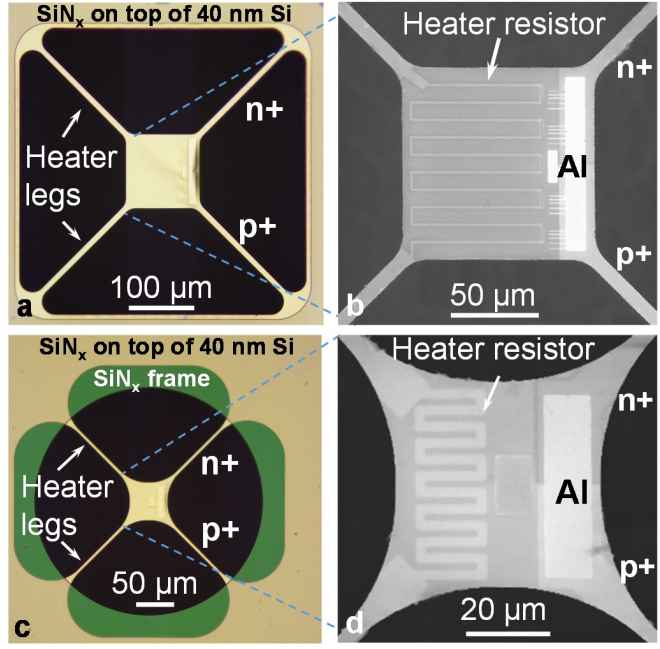

**FIG.1.** Optical and scanning electron microscope images of the bolometers: (a, b) device A and (c, d) device B. The pair of n- and p- heavily doped silicon beams shown on the right sides of the devices support the 40 nm thick silicon membrane and act as a thermocouple. The pair of the heavily doped silicon beams on the left side of the devices provides the electrical connection for the meandered doped silicon heater resistor patterned in the silicon membrane. The aluminum metal film forms a low-ohmic electrical contact between n- and p- doped thermocouple areas.

The lateral dimensions of the legs are lengths $l_n$ and $l_p$, and widths $w_n$ and $w_p$. In all other cases, $\widetilde{ZT} < ZT$ due to the contribution of extra heat links in the device and possibly due to additional series electrical resistance in the thermocouple circuit. We note also that $ZT$ reaches its maximum value $(ZT)_{max} = \frac{S^2 T}{\left(\sqrt{k_n \rho_n} + \sqrt{k_p \rho_p}\right)^2}$ in the optimal case, determined only via the thermoelectric material parameters, when the ratio of the number of squares for the p- ($N_p = l_p / w_p$) and n- legs ($N_n = l_n / w_n$) is $\frac{N_p}{N_n} = \sqrt{\frac{k_p \rho_n}{k_n \rho_p}}$. In the case of identical leg geometries $ZT = \frac{S^2 T}{(k_n + k_p)(\rho_n + \rho_p)}$, and in the case of similar material parameters for n- and p- legs ($k = k_n \approx k_p, \rho = \rho_n \approx \rho_p, S' = S_n \approx -S_p$) this further simplifies to the text book expression $ZT \approx (ZT)_{max} \approx S'^2 T / \rho k$.

The bolometers of this work were fabricated in the VTT Micronova clean room facilities on 150 mm silicon-on-insulator (SOI) wafers. The processing began with thinning of 270 nm thick SOI layer to the target thickness of 40 nm by thermal oxidation and oxide stripping. Next, the SOI layer were doped selectively with boron and phosphorus by ion implantation using photoresist as masks. After the doping, the SOI layer was patterned by UV-lithography and plasma etching and the implanted dopants were activated by annealing at 950 C deg.. Next, a 280 nm thick low stress $Si_xN_y$ layer was deposited by LPCVD and patterned in such a way that it formed a strain compensation frame in the spirit of Ref. [8] where the compensation is on a perimeter of un-patterned membranes. A 30 nm thick layer of Al was then sputtered and patterned by wet etching. Finally, the Si nano-membranes were released by a deep reactive etching through the silicon wafer from the back side and HF vapor etching of the buried oxide layer.

The two types of the bolometer devices, which we denote as device A and device B, with 40 nm thick Si membranes and beams, are shown in Fig. 1. Device A (B) consists of 110×110 µm$^2$ (50×50 µm$^2$) membranes suspended by four 190 µm (100 µm) long and 8 µm (3.5 µm) wide beams. The highly phosphorous- and boron-doped silicon beams form the thermocouple, which is connected on the membrane by the low-ohmic aluminum contact with contact resistance of 2.8 Ω (6.3 Ω) for device A (device B) as estimated from the measured test structures fabricated on the same chip. Another pair of beams forms the connection to the heavily doped meander heater resistor (resistance $R_h$) on the membrane for power response calibration of the bolometer and for reference thermometry as detailed below. In device A the heater is n-doped and in device B it is p-doped. Carrier densities (resistivities) of $4.8 \cdot 10^{19}$ cm$^{-3}$ (1.9 mΩ.cm) and $7.9 \cdot 10^{19}$ cm$^{-3}$ (1.8 mΩ.cm) for n- and p-type regions, respectively, were measured from van der Pauw test structures on the same chip. Note that in 40 nm thick single crystal silicon based membranes the thermal conductivity is reduced from the bulk value of 148 Wm$^{-1}$K$^{-1}$ down to ~ 30 Wm$^{-1}$K$^{-1}$ due to enhanced surface scattering of phonons [10]. In this thickness range the doping has smaller effect on the thermal conductivity [16]. The performance of the two bolometer structures was characterized in a vacuum chamber with the base pressure below 10$^{-2}$ Pa. The parameters for devices A and B are given in Table I. We note that $R$ is the total measured resistance including parasitic series connections on the chip, and the resistance of the free-standing part of the thermocouple is estimated to be lower based on the number of squares on the beams, yielding 23 kΩ and 27 kΩ for device A and B, respectively.

**TABLE I.** The parameters of bolometer devices A and B. $R$ is the total resistance of the thermocouple. $\tau_0$ is the thermal time constant and $dV/dP$ and **NEP** = $\sqrt{4k_B TR}$ /($dV$/$dP$) are the power responsivity and the total noise equivalent power at frequencies below the thermal cut-off **1/**$\tau_0$, respectively.

| Device | $R$ (kΩ) | $\tau_0$ **(ms)** | $dV/dP$ **(kV/W)** | **NEP (pW/Hz$^{1/2}$)** |
|---|---|---|---|---|
| A (110$^2$ μm$^2$) | 27 | 9.4 | 1.18 | 18 |
| B (50$^2$ μm$^2$) | 39 | 2.5 | 1.96 | 13 |

To measure the response of the bolometers, an AC voltage at frequency of $f$**/2** was applied to the heater resistor resulting in $f$-modulation of the power. The output voltage of the thermocouple was measured at frequency $f$ with lock-in techniques. The frequency responses of devices A and B are shown in Fig. 2a. The 3 dB cut-off thermal time constants $\tau_0$ and the responsivities $dV/dP = SG^{-1}$ (below the thermal cut-off) of the devices are obtained from the frequency response measurements by fitting the voltage amplitude data to $V = PSG^{-1}/\sqrt{1 + (2\pi f \tau_0)^2}$ where P is the amplitude of the power modulation (Fig. 2a). In the analysis it was taken into account that half of the Joule power generated in the beams contributes to the heating of the devices. The excellent agreement of the fits with the experimental data over the whole frequency range and with over 3 orders of magnitude in power amplitude demonstrates high linearity (see Fig. 2b) and good dynamic range of the devices. The fitting parameter values ($\tau_0$ and **/**$dP$ ) are given in Table I.

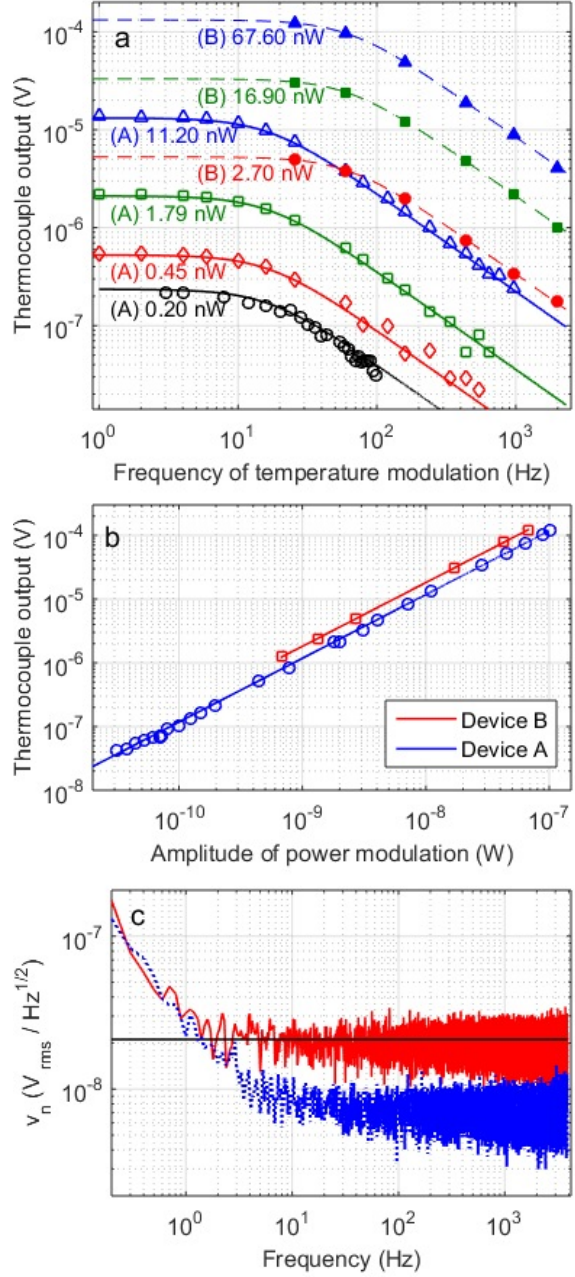

**FIG. 2.** (a) The frequency response of devices A and B. Symbols are the measured thermocouple output voltage amplitudes as function of the frequency of heater power at several heater power values. The curves show the response $V = PSG^{-1}/\sqrt{1 + (2\pi f \tau_0)^2}$ fitted to the data. The fitting parameters $\tau_0$ and $dV/dP = SG^{-1}$ are listed in Table I. (b) The measured amplitudes and the linear fits (lines) of thermocouple output voltages for devices A and B as function of the amplitude of the heater power modulation at $f = 4$ Hz (Device A) and $f = 26$ Hz (Device B). (c) The measured noise spectral density $v_n$ of the Device A (red curve) and the voltage noise floor of the measurement setup (blue dots), both referred to the preamplifier input. The black line shows the Johnson-Nyquist voltage noise density level of 21 nV/Hz$^{1/2}$ generated by a 27 kΩ resistor at 300 K.

The measured voltage noise spectrum (Fig. 2c) of the thermocouple referred to the preamplifier input was obtained by performing numerical FFT of the time-stamped voltage noise acquired with the bandwidth resolution of 0.1 Hz and averaged ten times to obtain better noise level estimation. The measured voltage noise level of $v_n \approx 21$ nV/Hz$^{1/2}$ corresponds to the Johnson-Nyquist noise-floor value $\sqrt{4k_B TR}$ of the thermocouple resistance $R = 27$ kΩ at 300 K. The measured reference noise level of the preamplifier is 8 nV/Hz$^{1/2}$ with its low frequency 1/f component emerging below ~ 3 Hz (see Fig. 2c). This shows that the noise in the present devices is dominated by the Johnson-Nyquist noise rather than phonon noise or readout noise, and hence indicates that $\widetilde{ZT}$ is significantly below unity. The measured level of the voltage noise allows us to calculate NEP of the devices. For the device A with total thermocouple resistance $R = 27$ kΩ and responsivity of 1.18 kV/W we find below thermal cut-off NEP = $v_n/(dV/dP) = 18$ pW/Hz$^{1/2}$. For device the B with the thermocouple resistance $R = 39$ kΩ the corresponding Johnson-noise limited NEP is 13 pW/Hz$^{1/2}$. Note that NEP for device B without the parasitic series connection would improve down to 11 pW/Hz$^{1/2}$.

To determine the total Seebeck coefficient $S$, the thermocouple of Device A was employed as a Peltier cooling and heating element by passing DC-current $I_{DC}$ through it. The resistance of the heater resistor $R_h(T_0)$ calibrated against bath temperature $T_0$ was used as a thermometer to determine the membrane temperature $T(I_{DC})$. The heater resistance was measured by lock-in techniques at 3 Hz with the excitation voltage of 10 mVrms, corresponding to the power bias of 0.45 nW, which is negligible compared to the Joule heating ($\propto I_{DC}^2$) and Peltier cooling/heating ($\propto I_{DC}$) power. Using Peltier coefficient of $T_0 S$ while requiring that all heat components of the system are in balance, the bolometer temperature can be written as $T - T_0 = aI_{DC}^2 - bI_{DC}$, where $a$ and $b$ are the fitting parameters. To ensure steady-state condition, i.e. that the membrane temperature is stabilized, the thermocouple DC current was swept up and down at a slow rate (3.7 nA/s). The absence of hysteresis in the data of Fig. 3 confirms that the device is in steady state. The total Seebeck coefficient can be calculated as $S = 0.5\, bR_{legs}/(aT_0)$, where the prefactor 0.5 is the fraction of the thermocouple Joule heat contributing to the membrane heating and $R_{legs} = 23$ kΩ is the resistance of the suspended thermocouple beams. The model fits very well to the experimental data, and the extracted value of $S = 387$ µV/K is in line with the sum of the absolute values of reported Seebeck coefficients of degenerately doped n- and p- type silicon [15, 17]. The measured $S$ and $dV/dP$ values yield thermal conductance $G = S(dV/dP)^{-1}$ of 0.33 µW/K and 0.20 µW/K, and the effective bolometer thermoelectric figures of merit $\widetilde{ZT}$ of 0.005 and 0.006 for devices A and B, respectively, from Eq. (1). Without the parasitic series resistance $\widetilde{ZT}$ increases to 0.006 and 0.008 for devices A and B, respectively.

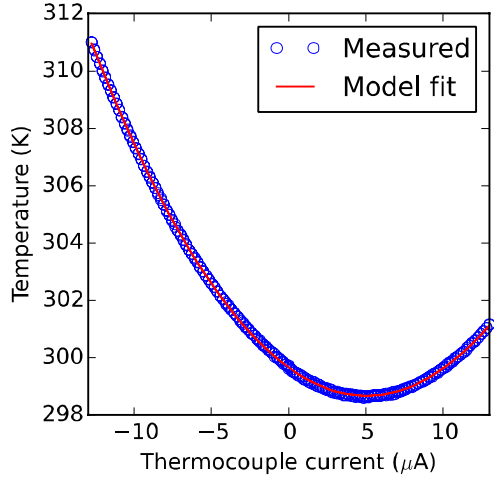

**FIG. 3.** The membrane temperature of Device A as a function of the DC current passed through the thermocouple (circles). The solid curve shows the parabolic fit of the heat balance model $T - T_0 = aI_{DC}^2 - bI_{DC}$. The fitted parameters are $a = 3.88 \cdot 10^{10}$ K/A$^2$, b = $3.91 \cdot 10^5$ K/A, and $T_0 = 299.7$ K.0.20 µW/K.

We now turn our discussion on how the performance of the devices can be further improved and discuss the feasibility for the IR detector applications. Although the realized devices are not optimized for optical detector applications, the obtained numbers, 18 pW/Hz$^{1/2}$ with $\tau_0 \sim$ 9.4 ms, and 13 pW/Hz$^{1/2}$ with $\tau_0 \sim$ 2.5 ms, are already competitive with the performance of thermopile detectors [18, 19, 20, 2] both in terms of power resolution and thermal speed. It should be noted that the devices need to be equipped with infrared absorbers which may increase the thermal mass and reduce the speed. However, thermal mass can be minimized by employing thin film IR absorbers [21]. Note that the two beams out of four in the devices are used for the heater wiring, which is not needed in an infrared detector. Removing these extra beams would halve $G$ to 100 nW/K and double the thermoelectric figure of merit $\widetilde{ZT}$ to 0.017 in the absence of any series resistance. A 2-beam version of device B modified for radiation detection would have an NEP of 5.5 pW/Hz$^{1/2}$ (given by Eq. 1) and a time constant around 5 ms due to the smaller $G$. A detector can also have two thermocouples connected in series. This kind of 4-beam version of device B would have the same $G$ and $\tau_0$ as before, and NEP of 7.8 pW/Hz$^{1/2}$. Further improvement in $\widetilde{ZT}$ can be achieved by scaling down the thickness of the Si membrane down to ~10 nm, which reduces the thermal conductivity of the Si membrane down to 9 W/m/K [10]. The 10 nm thick 2-beam version of device B would thus have G = 7.5 nW/K and $\widetilde{ZT}$ = 0.056 leading to NEP = 0.8 pW/Hz$^{1/2}$. This clearly shows the benefits of the scaled-down silicon thermoelectrics and, indeed, the most significant improvement on power resolution can be achieved by enhancing $\widetilde{ZT}$ up to ~1 by using, e.g., silicon nanowire beams [7, 6] with cross-sectional dimensions of 20 - 50 nm. Such an approach reduces the thermal conductivity of silicon by ~100-fold from its bulk value. This would push the bolometer resolution to the thermal fluctuation noise limited regime and sub-1 pW/Hz$^{1/2}$ NEP could be obtained with *G* below 100 nW/K without penalty in the thermal speed.

In summary, we have demonstrated nano-thermocouple bolometers based on ultra-thin heavily doped silicon membranes. The devices are micromachined essentially using only standard silicon and aluminum materials, making it highly attracting for the monolithic integration with CMOS readout circuits. The absence of additional micromechanical support layers enables near-material-limited operation without parasitic thermal conductivity and heat capacity, and enables full utilization of low dimensionality in thermoelectric performance optimization. The structure also provides a platform for reliable studies of the physics of low-dimensional thermoelectric transport, a subject of some controversy still [4, 22]. Operation with high thermal speed of 2.5 ms and low noise equivalent power of 13 pW/Hz$^{1/2}$ was demonstrated making the devices suitable for detection of small thermal signals. The performance of the bolometers is limited by the Johnson-Nyquist noise of the thermocouple, and by scaling down the size of the thermocouple beams to further reduce the phonon thermal conductance, thermal fluctuation noise limited performance can be potentially reached.

**ACKNOWLEDGMENTS**

This work has been financially supported by the European Union Seventh Framework Programme (grant agreement 604668, project QUANTIHEAT, and grant agreement 309150, project MERGING) and by the Academy of Finland (Grants No. 295329 and 252598 and the Finnish Centre of Excellence in Atomic Layer Deposition). Merja Markkanen is gratefully acknowledged for technical assistance in the sample fabrication.

**REFERENCES**

[1] F. Baldini, A. N. Chester, J. Homola, and S. Martellucci (Eds.), "Optical Chemical Sensors", Springer, 2006.

[2] U. Dillner, E. Kessler, and H.-G. Meyer, J. Sens. Sens. Syst. **2**, 85 (2013).

[3] A. Rogalski, Progress in Quantum Electronics **36**, 343 (2012).

[4] A. Shakouri, Annu. Rev. Mater. Res. **41**, 399 (2011).

[5] G. Penneli, Beilstein J. Nanotechnol. **5**, 1268 (2014).

[6] A. I. Boukai, Y. Bunimovich, J. Tahir-Kheli, J.-K. Yu, W. A. Goddard, and J.R. Heath, , Nature **451**, 168 (2008).

[7] A. I. Hochbaum, R. Chen, R. D. Delgado, W. Liang, E. C. Garnett, M. Najarian, A. Majumdar, and P. Yang, Nature **451**, 163 (2008).

[8] A. Shchepetov, M. Prunnila, F. Alzina, L. Schneider, J. Cuffe, H. Jiang, E. I. Kauppinen, C. M. Sotomayor-Torres, and J. Ahopelto, Appl. Phys. Let. **102**, 192108 (2013).

[9] B. Graczykowski, J. Gomis-Bresco, F. Alzina, J. S. Reparaz, A. Shchepetov, M. Prunnila, J. Ahopelto and C. M. Sotomayor Torres, New J. Phys. **16**, 073024 (2014).

[10] S. Neogi, J.S. Reparaz, L. F. C. Pereira, B. Graczykowski, M. R. Wagner, M. Sledzinska, A. Shchepetov, M. Prunnila, J. Ahopelto, C. M. Sotomayor-Torres, and D. Donadio, ACS Nano **9**, 3820 (2015).

[11] S. Gomès, A. Assy, and P-O. Chapuis, Phys. Stat. Sol. A **212**, 477 (2015).

[12] K. Kim, B. Song, V. Fernández-Hurtado, W. Lee, W. Jeong, L. Cui, D. Thompson, J. Feist, M. T. Homer Reid, F. J. García-Vidal, J. C. Cuevas, E. Meyhofer, and P. Reddy, Nature **528**, 387 (2015).

[13] N. Neophytou, X. Zianni, H. Kosina, Stefano Frabboni, B. Lorenzi, and D. Narducci, Nanotechnology **24**, 205402 (2013).

[14] N. Neophytou, H. Karamitaheri, and H. Kosina, J. Comput. Electron. **12**, 611 (2013).

[15] H. Ikeda and F. Salleh, Appl. Phys. Lett. **96**, 012106 (2010).

[16] M. Asheghi, K. Kurabayashi, R. Kasnavi, K.E. Goodson, J. Appl. Phys. **91**, 8 (2002).

[17] A. Stranz, J. Kähler, A. Waag, and E. Peiner, J. Electronic Materials **42**, 2381 (2013).

[18] M. C. Foote, M. Kenyon, T. R. Krueger, T.A. McCann, R. Chacon, E.W. Jones, M.R. Dickie, J.T. Schofield, and D.J. McCleese, Proc. of the International Workshop on Thermal Detectors for Space Based Planetary, Solar, and Earth Science Applications, Adelphi, USA, June 2003, 2, 16-20, (2003).

[19] F. Haenschke, E. Kessler, U. Dillner, A. Ihring, U. Schinkel, and H.-G. Meyer, Infrared Tech. and Appl. XXXVIII, Proc. of SPIE Vol. 8353, 83531L (2012).

[20] M. Hirota, Y. Nakajima, M. Saito, and M. Uchiyama, Sensors and Actuators A **135**, 146 (2007).

[21] J. J. Talghader, A. S. Gawarikar, and R. P. Shea, Light: Science & Applications **1**, e24 (2012).

[22] O. Caballero-Calero, and M. Martin-González, Scripta Materialia **111**, 54 (2016).